\documentclass [10pt]{iopart}
\usepackage{epsf}
\usepackage{multirow}
\usepackage{graphics}
\eqnobysec
\newcommand{\rem}[1]{\hbox{}}
\begin{document}
\title{Cylindrical Spacetimes with $\Lambda \not =0$ and their Sources}
\author{M. \v{Z}ofka\dag, J. Bi\v{c}\'{a}k\dag\ddag}
\address{\dag Institute of Theoretical Physics, Faculty of Mathematics and Physics, Charles University Prague, V Hole\v{s}ovi\v{c}k\'{a}ch 2, 180 00 Prague 8, Czech Republic\\
\ddag Albert-Einstein-Institute, Max-Planck-Institute for Gravitational Physics, Golm, Germany\\}
\eads{\mailto{bicak@mbox.troja.mff.cuni.cz}\\\mailto{zofka@mbox.troja.mff.cuni.cz}}
\date{}
\begin{abstract}
We review and investigate some basic properties of static, cylin\-drically symmetric spacetimes with non-zero cosmological constant, find non-singular sheet sources of these spacetimes and discuss their characteristics, and clarify their relation to the 4D black-string solutions.
\end{abstract}
\submitto{\CQG}
\section{Introduction}
The present paper is a continuation of our previous work on shell sources of the Levi-Civita spacetime (LC) \cite{BiZo}. The thin-shell formalism \cite{Israel} provides insight into the physical meaning of the parameters appearing in the metric and can set limits on their admissible ranges. Our aim is to extend the studied spacetimes to include a non-zero cosmological constant $\Lambda$. In the present-day cosmology, the role of spacetimes with $\Lambda>0$ has grown significantly since models with a positive effective cosmological constant describe the inflationary phase and, moreover, recent observations still seem to indicate that our universe indeed has $\Lambda>0$. On the other hand, the anti de Sitter spacetime (AdS) has come to the fore from a more theoretical perspective in light of the conjecture relating string theory in asymptotically AdS space to a non-gravitational conformal theory on the boundary at spatial infinity. However, the motivation for considering material systems with cylindrical symmetry in asymptotically AdS spacetimes comes also from the `classical general relativity'. It is well known that with $\Lambda=0$, the asymptotics of cylindrically symmetric static spacetimes representing {\em infinite} sources is very different from that of spatially {\em bounded} static sources. As we shall indicate below, with $\Lambda<0$, the asymptotic forms of the metrics due to material cylinders and bounded sources are closely related. It is therefore of interest to study the effects of both positive and negative cosmological constants on the matter sources of cylindrical spacetimes. At present, we restrict ourselves to static spacetimes since the involved effects are still rich enough while the number of free parameters is lower and thus it is easier to interpret our results. In general, we admit spacetimes outside and inside the shell to have different values of $\Lambda$---if the cosmological constant is considered to be vacuum energy, such assumption is not unnatural (and, indeed, has been made in literature in case of spherical shells giving rise to the so-called domain walls \cite{Domain_walls}). We can also introduce flat or AdS spacetime inside the cylinder. Although in this work we only consider static metrics and their sources, we also realize their possible applications as background spacetimes in the study of gravitational waves produced by material sources---specific simple perturbations of cylindrical spacetimes allow a rather explicit treatment \cite{Bondi}.

The line element of a static, cylindrically symmetric spacetime with a non-zero cosmological constant (LC$\Lambda$) was independently derived in \cite{Linet} and \cite{Tian}. In cylindrical coordinates $t,z \in \! I\hspace{-0.13cm}R,r \in \! I\hspace{-0.13cm}R^+, \varphi \in [0,2\pi)$ (the standard range of $\varphi$ is ensured by the presence of the conicity parameter $C$) the metric reads
\begin{eqnarray}\label{LC+Lambda}
\hspace{-1cm}ds^2  = && Q(r)^{2/3} \{ -P(r)^{-2(4\sigma^2-8\sigma+1)/3A} dt^2 
+ P(r)^{2(8\sigma^2-4\sigma-1)/3A} dz^2 + \\
&& + P(r)^{-4(2\sigma^2+2\sigma-1)/3A} d\varphi^2/C^2 \} +dr^2, \nonumber
\end{eqnarray}
where
\begin{equation}\label{Definition of A using sigma}
A \equiv 4\sigma^2-2\sigma+1,
\end{equation}
$\sigma$ is related to the linear mass density of the source, and for $\Lambda>0$ we have
\begin{equation}\label{Metric, positive Lambda}
P(r)=\frac{2\tan (\sqrt{3\Lambda}r/2)}{\sqrt{3\Lambda}}, \:\: Q(r)= \frac{\sin (\sqrt{3\Lambda}r)}{\sqrt{3\Lambda}},
\end{equation}
while for $\Lambda<0$ we substitute trigonometric functions by their hyperbolic counterparts and $\Lambda$ by $-\Lambda$---see \ref{Derivation of the Metric}. The trigonometric functions in (\ref{Metric, positive Lambda}) are multiplied by such constants that in the limit $r \to 0$ or $\Lambda \to 0$ the metric smoothly approaches the LC case ($\Lambda=0$). Thus, the dimensions of $t,z,$ and $1/C$ are general powers of length as is the case in the standard form of the LC metric \cite{Kramer}. By redefining $P(r)$ and $Q(r)$ we can convert coordinates to the usual dimensions of length.

\label{Discussion of asymptotics}The advantage of the studied spacetimes over the LC case is that for $\Lambda<0$, far away from the symmetry axis, $r=0$, they approach the anti de Sitter solution (AdS) unlike LC that does not approach the Minkowski spacetime. We show explicitly that if we transform a Schwarzschild-anti de Sitter black hole (SAdS) from Schwarzschild to horospherical coordinates then far away from $r=0$, the leading-order diagonal terms are identical to the LC$\Lambda$ expansion (and to AdS) and the second-order terms have the same dependance on the radial coordinate, $r$, which is to the first order the proper distance from the axis. This indicates that the properties of cylindrical shells with $\Lambda<0$ might be of importance in the study of finite massive objects in contrast to the asymptotically Minkowskian spacetimes and the LC solution.

Moreover, Lemos \cite{Lemos} found an interesting class of cylindrically symmetric 4D solutions with $\Lambda<0$ called black strings.\footnote{Not to be confused with higher-dimensional black strings that are direct products of the Schwarzschild spacetime with $I\hspace{-0.13cm}R$ (see Section 5 in \cite{Horowitz}).} These spacetimes describe the fields of charged, rotating strings that generally feature singularities and horizons and are asymptotically anti de Sitter far away from the axis of symmetry. We clarify here the relation of static, uncharged black strings to the general LC$\Lambda$ solutions and find non-singular sources of such spacetimes.

The paper is organized as follows: In Section 2, we discuss various properties of LC$\Lambda$ spacetimes, in particular, their relation to the black-string metrics and their asymptotic form, we introduce suitable quantities characterizing their spatial geometry and conicity (we also correct some minor misprints in the literature). The shell sources of LC$\Lambda$ are constructed in Section 3 where we analyze the physical properties of matter forming the shells and find the resulting restrictions on the LC$\Lambda$ parameters. We relegate a number of issues into 3 appendices. In \ref{Derivation of the Metric}, we derive the LC$\Lambda$ metric following \cite{Tian} and point out several interesting details such as its precise relation to the AdS spacetime. \ref{Asymptotics} gives the resulting form of the asymptotic expansion of (\ref{LC+Lambda}) for $\Lambda<0$ and compares it to the SAdS spacetime. Finally, in \ref{Bound on M1}, we discuss the admissible range of the mass per unit length for shell sources of the LC$\Lambda$ spacetimes.
\section{Properties of the Metric}
It is important to realize that the LC$\Lambda$ solutions (\ref{LC+Lambda}) are not conformally flat unless we put $\Lambda=0$---and thus also not conformal to (anti) de Sitter spacetime---since the Weyl tensor is non-zero. In general, the solutions are of Petrov type I except for some special values of $\sigma$ \cite{da_Silva_et_al}. In addition to the Killing vectors $\partial / \partial t,\partial / \partial z$, and $\partial / \partial \varphi$ the metrics admit additional Killing vectors in special cases\footnote{There are misprints in the Killing vectors in \cite{da_Silva_et_al}. Also, the presented conformal diagrams do not have correct infinities---compare to \cite{Lemos}.}:
\begin{eqnarray}\label{Killing vectors}
\sigma = \pm 1/2 & \rightarrow & C^{-1} \varphi \partial z - Cz \partial \varphi, \nonumber\\
\sigma = 0,1 & \rightarrow & t \partial z + z \partial t,\\
\sigma = 1/4 & \rightarrow & C^{-1} \varphi \partial t + Ct \partial \varphi.\nonumber
\end{eqnarray}
In the vicinity of the axis, $r\sqrt{|\Lambda|} \ll 1$, (\ref{LC+Lambda}) behaves as the LC solution. After replacing $\sigma \rightarrow 1/4 \sigma$, we find the roles of $z$ and $\varphi$ interchanged. This indicates that the case $\sigma= \pm 1/2$ corresponds to planar rather than cylindrical symmetry (confer the first Killing vector in (\ref{Killing vectors})).

The solution with $\Lambda>0$ is periodic in $r$ with a period $2\pi/\sqrt{3\Lambda}$; it contains curvature singularities at $r = k\pi/\sqrt{3\Lambda} \equiv kr_s, k$ integer except for special cases of $\sigma$ \cite{da_Silva_et_al}; there is no limit $r \rightarrow \infty$ and the metric is {\it not} asymptotically de Sitter.

For $\Lambda<0$, there are no singularities apart from the axis. After a simple rescaling of $t,z,\varphi$ we find that asymptotically at $r \rightarrow \infty$, the metric (\ref{LC+Lambda}) has the form
\begin{equation}\label{Horospherical coordinates}
ds^2 = dr^2 + \exp (2 \; \sqrt{-\frac {\Lambda} {3}} \; r) \left( -dt^2+dz^2+d\varphi^2 \right),
\end{equation}
which is a part of AdS in horospherical coordinates (see \ref{Derivation of the Metric}).  This makes the asymptotics of cylindrical solutions with $\Lambda<0$ more realistic than the case $\Lambda=0$. \label{Discussion of asymptotics part II}If we approach infinity in any direction not parallel to the axis then the leading terms of the asymptotics of the metric coincide with those of spatially isolated sources. For $\Lambda = 0$ this is not the case since the asymptotics of the LC solution is {\it not} Minkowskian and we have no natural (asymptotically flat) infinity that would enable us to define static observers. To illustrate this, we compare the asymptotics of (\ref{LC+Lambda}) and that of Schwarzschild-AdS spacetime in horospherical coordinates in \ref{Asymptotics}.

It is of interest that LC$\Lambda$ with $\Lambda<0$ includes the static, uncharged black-string metric of \cite{Lemos}. If we set $\sigma=1/2$ in (\ref{LC+Lambda}) and introduce new coordinates
\begin{equation}
\tau = \frac {2 C}{(-\Lambda)} \, t, \;\;\; \rho = \frac{1}{C} \cosh^{2/3} \frac {\sqrt{-3\Lambda}r} {2}, \;\;\; \phi = \varphi, \;\;\; \zeta = \sqrt{-\frac {3} {\Lambda}} C z,
\end{equation}
we obtain:
\begin{equation}\label{Black-string Metric}
ds^2 = -d\tau^2 \mathcal{F} + \frac{d\rho^2}{\mathcal{F}} + \rho^2 d\phi^2 + \rho^2 \alpha^2 d\zeta^2,
\end{equation}
with $\alpha^2 = (-\Lambda)/3$, $\mathcal{F}= \alpha^2 \rho^2 - 4M/\alpha\rho$, and $M=\alpha^3/4C^3$. This is the form of the black-string metric of Equation (2.3) in \cite{Lemos}. $M$ corresponds to the mass per unit length of the black string calculated at radial infinity using the Hamiltonian formalism of Brown and York \cite{Brown_and_York}. Therefore, it is instructive to compare $M$ to the mass per unit length of the shell sources we find in Section \ref{Induced Energy-momentum Tensor}.

To illustrate some other properties of LC$\Lambda$, we summarize the circumferences $\mathcal{C}$ of the rings $r,z,t \! = \:$constant at special radii $r$ in \Tref{Circumferences of Rings}.
\begin{table}[t]
$$
\begin{array}{||c|c|c|c|c||}
\hline \hline
\mbox{Sign of}          & \mbox{Value of}   & \mbox{Value of}   & \mbox{Circumference of} & \mbox{Singularity?} \\
\hspace{0.5cm} \Lambda  & r                 & \sigma             & \mbox{rings} \; r, z, t \; \mbox{constant} & \mbox{(Kretschmann)}\\
\hline
\hline
\multirow{6}{*}{$\Lambda=0$}   && \sigma < 1/2 & \mathcal{C}=0 & \mbox{Yes, for } \sigma \not=0 \\
            & r=0 & \sigma = 1/2 & \mathcal{C} \mbox{ finite} & \mbox{No} \\
            && \sigma > 1/2 & \mathcal{C} \rightarrow \infty & \mbox{Yes}\\
\cline{2-5}
            && \sigma < 1/2 & \mathcal{C} \rightarrow \infty & \mbox{No} \\
            & r \rightarrow \infty & \sigma = 1/2 & \mathcal{C} \mbox{ finite} & \mbox{No} \\
            && \sigma > 1/2 & \mathcal{C} =0 & \mbox{No}\\
\hline
\hline
\multirow{2}{*}{$\Lambda<0$}
            & r=0 & \multicolumn{2}{c}{\mbox{Same as for } \Lambda=0, r=0} & \\
\cline{2-5}
            & r \rightarrow \infty & \mbox{Arbitrary} \;\; \sigma & \mathcal{C} \rightarrow \infty & \mbox{No} \\
\hline
\hline
\multirow{6}{*}{$\Lambda>0$}   & r = 2k \pi/\sqrt {3 \Lambda}             & \multicolumn{2}{c}{\mbox{Same as for } \Lambda=0, r=0} &\\
\cline{2-5}
            && \sigma < -1/2 & \mathcal{C}=0 & \mbox{Yes}\\
            && \sigma = -1/2 & \mathcal{C} \mbox{ finite} & \mbox{No} \\
            & r = \frac {(2k+1)\pi} {\sqrt {3 \Lambda}} & \sigma \in (-1/2,1/4) & \mathcal{C} \rightarrow \infty & \mbox{Yes}\\
            && \sigma = 1/4 & \mathcal{C} \mbox{ finite} & \mbox{No} \\
            && \sigma > 1/4 & \mathcal{C}=0 & \mbox{Yes, for } \sigma \not= 1 \\
\hline \hline
\end{array}
$$
\caption{\label{Circumferences of Rings} Geometrical properties of LC and LC$\Lambda$. Circumferences of rings $r,z,t = \; $constant and curvature singularities at $r=0$ and $r \rightarrow \infty$ in the cases $\Lambda =0$ and $\Lambda<0$, and at $r=k\pi / \sqrt{3\Lambda}$ for $\Lambda>0$. In the following we shall see that physically meaningful shell sources generate spacetimes with $\sigma$ confined only to the interval $[0,1/2]$.}
\end{table}
For example, with $\Lambda>0$ and $\sigma \in (-\infty,-1/2 \, ] \cup [1/4,1/2 \, ]$ the circumference remains finite everywhere, including $r=0$. On the other hand, $r=0$ represents the axis (zero circumference) for $\sigma<1/2$ and any value of $\Lambda$. If we also require that $r=r_s$ has an infinite circumference for $\Lambda>0$ then the range of $\sigma$ is more limited: $\sigma \in (-1/2,1/4)$. For $\Lambda<0$, the circumference is always infinite at $r \rightarrow \infty$ and we find $\sigma<1/2$ again.

In order to characterize the conicity of cylindrically symmetric spacetimes in a geometrical way, we define the following two quantities
\begin{eqnarray}\label{Conicity definition}
\chi(r) & \equiv & 2 \pi \frac {R(r)} {\mathcal{C}(r)} = \frac {\int _0^r \sqrt {g_{rr} (\tilde{r})} d\tilde{r}} {\sqrt {g_{ \varphi \varphi} (r)}} = \frac {2 \pi r} {\mathcal{C}(r)},\\
\psi(r) & \equiv & 2\pi \frac {dR} {d\mathcal{C}} = 2 \frac {\sqrt {g_{ \varphi \varphi} (r) g_{rr} (r)}} {\frac {d} {dr} g_{ \varphi \varphi} (r)} = \frac{1} {\frac {d} {dr} \sqrt{g_{ \varphi \varphi}(r)}},
\end{eqnarray}
where $R$ is the proper radius (in our case identical to $r$) and the circumference $\mathcal{C}$ is introduced above. In case of cosmic strings with $\Lambda = \sigma = 0$, both $\chi$ and $\psi$ coincide everywhere with the conicity parameter $C$. Close to the axis, these quantities behave as follows
\begin{equation}\label{Psi}
\chi(r) \sim C r^{\frac {4 \sigma ^2}{A}} \; \mbox{ and } \; \psi(r) \sim \frac {A} {1-2 \sigma} \; C r ^{\frac {4 \sigma ^2} {A}},
\end{equation}
where $A>0$ is defined in (\ref{Definition of A using sigma}). They do not contain $\Lambda$ and are identical to the Levi-Civita case. In general, $\psi$ can even be negative, nevertheless, for $\Lambda>0, \sigma \in (-1/2,1/4)$ and for $\Lambda<0, \sigma<1/2$, $\psi$ is positive everywhere. For $\Lambda>0$, we find that as we approach the singularity at $r=r_s$ the value of $\psi$ is proportional to $(r_s - r)^{4( 1- \sigma)^2/3A}$ and thus it vanishes unless $\sigma=1$. For $\Lambda<0$, we find at radial infinity that $\chi$ and $\psi$ are both positive and decrease as a power of $\Lambda$ times $r \exp(-r \sqrt{\Lambda/3})$ and $\exp(-r \sqrt{\Lambda/3})$, respectively.

Considering the above properties, we conclude that the intuitive ranges of $\sigma$ are as follows: with $\Lambda>0, \sigma \in (-1/2,1/4)$ and with $\Lambda<0, \sigma \in (-\infty,1/2)$. If we also include the behavior of geodesics we can further restrict the latter to $\sigma \in [-1/2,1/2)$.
\section{Induced Energy-momentum Tensor}\label{Induced Energy-momentum Tensor}
To construct shell sources of LC$\Lambda$, we imagine two cylindrical regions of the spacetimes with metrics (\ref{LC+Lambda}) and independent parameters $(r \leq r_-,\sigma_-,\Lambda_-,C_-)$ and $(r \geq r_+,\sigma_+,\Lambda_+,C_+)$.  We shall also consider Minkowski and AdS solutions as the interior spacetimes. As required by the Israel formalism \cite{Israel}, we glue the boundaries $r=r_-$ and $r=r_+$ together in such a way that the induced 3-metric on the surfaces with intrinsic coordinates $T, Z, \mathit{\Phi}$ is identical from both sides; the induced metric can always be made flat by rescaling the intrinsic coordinates as the metric only depends on the radial coordinate $r$. We also require the proper circumference of the two cylinders to be the same. This identification induces a 3-dimensional energy-momentum tensor on the junction hypersurface. The induced tensor is diagonal and its components read\footnote{Unless noted otherwise, we write all the following expressions for $\Lambda>0$. The corresponding formulae for $\Lambda<0$ are obtained by substituting hyperbolic functions for their trigonometric counterparts and replacing $\Lambda$ by $-\Lambda$.}:
\begin{equation}\label{Induced Tensor with Lambda}\fl
\begin{array}{rcl}
8\pi S_{TT} & = & \sqrt {\Lambda_+/3} \left( -2\cot (\sqrt {3 \Lambda_+} \; r_+) - \frac{4\sigma_+^2 -8\sigma_+ +1}{4\sigma_+^2 -2\sigma_+ +1} \csc (\sqrt {3 \Lambda_+} \; r_+) \right) - \\
&& - \sqrt {\Lambda_-/3} \left( -2\cot (\sqrt {3 \Lambda_-} \; r_-)- \frac{4\sigma_-^2 -8\sigma_- +1}{4\sigma_-^2 -2\sigma_- +1} \csc (\sqrt {3 \Lambda_-} \; r_-) \right), \\
\\
8\pi S_{ZZ} & = & \sqrt {\Lambda_+/3} \left[ 2\cot (\sqrt {3 \Lambda_+} \; r_+)- \frac{8\sigma_+^2 -4\sigma_+ -1}{4\sigma_+^2 -2\sigma_+ +1} \csc (\sqrt {3 \Lambda_+} \; r_+) \right] - \nonumber \\
&& - \sqrt {\Lambda_-/3} \left[ 2\cot (\sqrt {3 \Lambda_-} \; r_-)- \frac{8\sigma_-^2 -4\sigma_- -1}{4\sigma_-^2 -2\sigma_- +1} \csc (\sqrt {3 \Lambda_-} \; r_-)\right],\\
\\
8\pi S_{\mathit{\Phi} \mathit{\Phi}} & = & \sqrt {\Lambda_+/3} \left( 2\cot (\sqrt {3 \Lambda_+} \; r_+)+ \frac{4\sigma_+^2 +4\sigma_+ -2}{4\sigma_+^2 -2\sigma_+ +1} \csc (\sqrt {3 \Lambda_+} \; r_+) \right) - \nonumber \\
&& - \sqrt {\Lambda_-/3} \left( 2\cot (\sqrt {3 \Lambda_-} \; r_-)+ \frac{4\sigma_-^2 +4\sigma_- -2}{4\sigma_-^2 -2\sigma_- +1} \csc (\sqrt {3 \Lambda_-} \; r_-) \right),
\end{array}
\end{equation}
\begin{flushleft}
with $A_{\pm}(\sigma_{\pm})$ defined by (\ref{Definition of A using sigma}). Note that the (global) conicity parameters $C_\pm$ of the metric do not appear in these (local) expressions. If we take the limit $\Lambda \rightarrow 0$ and use the transformation formulae given in \cite{da_Silva_et_al} we recover precisely the induced energy-momentum tensor of the Levi-Civita metric \cite{BiZo}, Equation 3.\end{flushleft}

\subsection{Photonic Shells}
Let us turn to the interpretation of the induced energy-momentum tensor. One of the simplest models is a shell of counter-rotating photons. Then the trace $-S_{TT} + S_{\mathit{\Phi} \mathit{\Phi}} + S_{ZZ}=0$ which implies
\begin{equation}\label{Photons with Lambda}
\sqrt{ \frac {\Lambda_+} {\Lambda_-} } = \frac {\tan (\sqrt
{3\Lambda_+} r_+)} {\tan (\sqrt {3\Lambda_-} r_-)}.
\end{equation}
Note that this condition does not involve $\sigma_\pm$. If, in addition, we require $\Lambda_+= \Lambda_- \equiv \Lambda>0$ we obtain the solutions $r_+ = r_- + k\pi / \sqrt{3 \Lambda}$ with an arbitrary integer $k$. These solutions are due to the periodicity of the metric and are difficult to interpret: We either have singularities outside the shell or, alternatively, we have to construct an infinite series of coaxial cylinders to cut out all the singularities. However, for a \emph{negative} cosmological constant, we only have a single solution, namely $r_+=r_-$, which can be non-singular both in- and outside of the shell.

Admitting still both signs of $\Lambda$, we now wish to find the form of the induced energy-momentum tensor that can be interpreted as a shell of photons counter-rotating in the $\Phi$-direction with $Z$ constant (`azimuthal' photons) or, possibly, moving along the $Z$-direction with $\Phi$ constant (`axial' photons). Since we wish to avoid the singularity along the axis we must set $\sigma_-=0$ or $1/2$. Substituting into (\ref{Induced Tensor with Lambda}), denoting the mass per unit length of the shell measured by static (geodetical) observers $M_1 \equiv \mathcal{C} S_{TT}$ (with $\mathcal{C}$ the circumference of the shell and $S_{TT}$ its surface energy density), and with $\mathcal{M} = \frac{1} {4} \frac {1} {\cos^{4/3}( \sqrt {3\Lambda}r_-/2)}$, we summarize the results in the following \Tref{Trivial Photons}:
\begin{table}[h]\label{Trivial Photons}
$$
\begin{array}{||c|c|c|c|c||}
\hline \hline
\multirow{4}{*}{$\sigma_-=0$} & \multirow{2}{*}{$k$ odd} & \sigma_+ = -1/2 & M_1 = \mathcal{M} & \mbox{azimuthal}\\
            && \sigma_+ = 1 & M_1 = 0 & \mbox{smooth junction}\\
\cline{2-5}
            & \multirow{2}{*}{$k$ even} & \sigma_+ = 1/2 & M_1 = \mathcal{M} & \mbox{azimuthal}\\
            && \sigma_+ = 0 & M_1 = 0 & \mbox{smooth junction}\\
\hline
\hline
\multirow{3}{*}{$\sigma_-= \frac{1}{2}$}  & \multirow{2}{*}{$k$ odd} & \sigma_+ = 1 & M_1 = -\mathcal{M}<0 & 
\mbox{azimuthal}\\
            && \sigma_+ = 1/4 & M_1 = -\mathcal{M}<0 & \mbox{axial}\\
\cline{2-5}
            & k \mbox{ even} & \sigma_+ = 0 & M_1 = -\mathcal{M}<0 & \mbox{azimuthal}\\
\hline \hline
\end{array}
$$
\caption{Shells of photons moving with ${Z=}$ const or ${\mathit{\Phi}=}$ const and no singularity along the axis ($\mathcal{M} = 1/4\cos^{4/3}( \sqrt {3\Lambda}r_-/2)$).}
\end{table}

For $\Lambda>0$, although some cylindrical shells are `physical' ($M_1>0$), the exterior metric always contains singularities. In the case $\sigma_+ = -1/2$, we need to extend the solution through a coordinate singularity. However, this is not the case with $\Lambda<0$, which is also described in \Tref{Trivial Photons} if we formally only consider $k$ even and put $\mathcal{M} = \frac{1} {4} \frac {1} {\cosh^{4/3}( \sqrt {-3\Lambda}r_-/2)} \leq \frac{1}{4}$. Clearly, the only physically acceptable solutions are those with $\Lambda<0, M_1>0$, so that $\sigma_-=0$, $\sigma_+=1/2$, which corresponds to cylinders of azimuthal photons. In the limit $\Lambda \rightarrow 0$ these results go over to the LC case of \cite{BiZo}, Section 3.

As in the simpler $\Lambda=0$ case, there are no cylinders of purely axial photons. One expects this for $\Lambda<0$ due to its attractive character. However, with $\Lambda>0$, one might have hoped that the repulsive effects of the cosmological constant might permit such a solution; this was our original motivation. Nevertheless, the singularities located at $r=k r_s$ do not allow this.

\subsection{Shells of Dust and Perfect Fluid}
Since we wish to discuss primarily smooth and physical metrics (\ref{LC+Lambda}) we shall be mainly concerned with LC$\Lambda$ spacetimes with $\Lambda<0$. Only in these cases we do not encounter singularities at $r=k r_s$. We first assume $\Lambda_-=\Lambda_+ \equiv \Lambda<0$ and put $\sigma_-=0, C_-=1$ so that there is no singularity on the axis; we shall consider also Minkowski and AdS regions inside the shells---then the axis is regular automatically. Let us first notice that condition (\ref{Photons with Lambda}) is also relevant to the massive-particle
interpretation. In this case we require
\begin{equation}
\label{particle_condition} S_{\mathit{\Phi} \mathit{\Phi}}/S_{TT} + S_{ZZ}/S_{TT} = v_{\mathit{\Phi}}^2 + v_{Z}^2 < 1.
\end{equation}
With $\Lambda<0$ we obtain a simple condition $r_+>r_-$. (This condition becomes more involved for $\Lambda>0$.)

\begin{figure}[h]
\begin{center}
\epsfxsize=2in 
\epsfbox{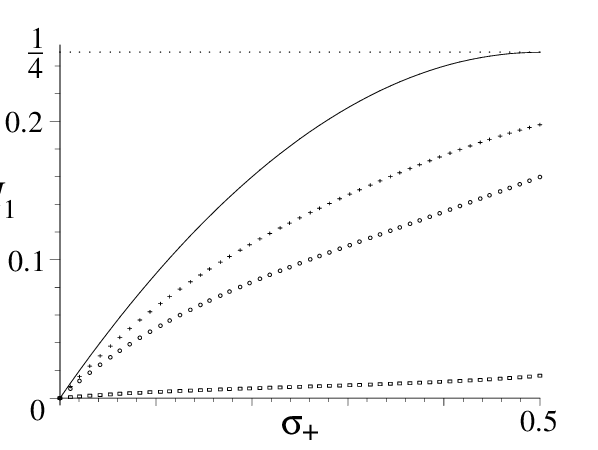}
\hspace{1cm}
\epsfxsize=2in
\epsfbox{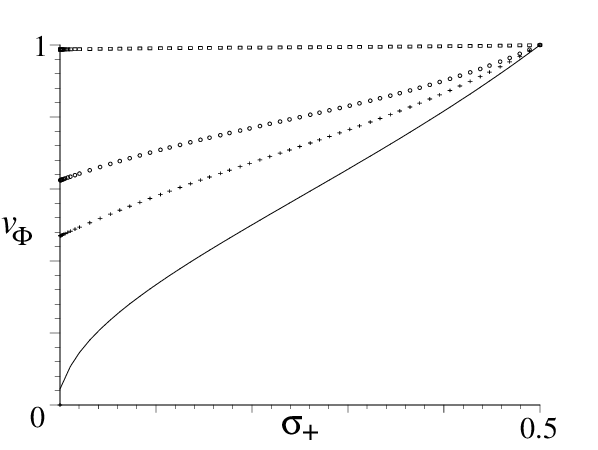}
\end{center}
\caption{\label{Azimuthal Particles} Mass per unit length (left) and proper velocity (right) for counter-rotating particles with $\sigma_-=0, \Lambda_-=\Lambda_+ \equiv \Lambda<0, C_-=1$. The plotted curves correspond to $\Lambda=-10,-1,-1/2,-1/1000$ (bottom to top in the left and top to bottom in the right) for a shell of fixed circumference. For $\sigma_+ \in [0,1/2)$ the proper velocity is finite and positive. The graph corresponding to $\Lambda=-1/1000$ (full curve) approaches the graph for a LC shell, cf. Fig 2 of \cite{BiZo}.}
\end{figure}

We begin by considering counter-rotating azimuthal streams of massive particles. The condition $S_{ZZ}=0$ implies (cf. (\ref{Induced Tensor with Lambda}))
\begin{equation}\label{Outer_radius}
r_+ = \frac{1}{\sqrt{-3\Lambda}} \; \mbox{arccosh} \frac {2E+F \sqrt {F^2+E^2-4}} {F^2
-4},
\end{equation}
where $E= (-8 \sigma_+^2 + 4 \sigma_+ +1) / (4 \sigma_+^2 -2
\sigma_+ +1)$ and $F= (2 \cosh (\sqrt {-3\Lambda} r_-) +1) / (\sinh (\sqrt {-3\Lambda} r_-))$. Fixing $r_-$, i.e., the circumference $\mathcal{C}$, we obtain the mass per unit length of the cylinder in the form
\begin{equation}\fl
\label{mass_per_unit_length} \hspace{1.5cm} M_1= \mathcal{C} S_{TT}  = \frac{2 \sigma_+ (1-\sigma_+) \sinh( \frac {1}{2} \sqrt {-3 \Lambda}r_-)} {\cosh ^{1/3} ( \frac {1}{2} \sqrt {-3 \Lambda}r_-)\sinh( \sqrt {-3\Lambda}r_+)(4 \sigma_+^2-2 \sigma_+ +1)}.
\end{equation}
Both $M_1$ and the velocity of particles counter-orbiting in the $\pm\varphi$ direction measured by static observers within the cylinder (see (\ref{particle_condition})) can be expressed as functions of $\sigma_+$ after substituting for $S_{TT}$ and $S_{FF}$ from (\ref{Induced Tensor with Lambda}). The final result is simple in the limit $\Lambda \rightarrow 0$ when the azimuthal velocity becomes $v_{\mathit{\Phi}}= \sqrt{\sigma_+/(1-\sigma_+)}$. This is the result obtained in \cite{BiZo}. We illustrate both functions in \Fref{Azimuthal Particles}. The physical properties are described in the figure caption, here we only notice that the velocity of particles approaches the velocity of light for $\sigma_+ \rightarrow 1/2$ thus allowing the interpretation in terms of massive counter-orbiting particles only for $\sigma \in [0,1/2]$. Also, observe that in general $M_1=0$ for $\sigma_+=0$ and that the curves $M_1(\sigma_+)$ are monotonically increasing for all values $\Lambda<0$ and $\sigma_+ \leq 1/2$.

\label{Discussion of Limits on M1} Given the results for the Levi-Civita spacetimes with $\Lambda=0$, we expect that there is an upper bound on $M_1$.  This, indeed, can be seen from Figure 1: the mass per unit length $M_1$ reaches its maximum value of 1/4 just for the LC case $\Lambda=0$. As $|\Lambda|$ increases the maximum of $M_1$ (located always at $\sigma_+=1/2$) decreases. This behavior can be proved analytically from (\ref{mass_per_unit_length}). It is physically intuitive in view of increased gravity due to the presence of the negative cosmological constant. A more detailed analysis reveals that the upper bound on $M_1=1/4$ holds also in more general cases with different $\Lambda$'s if $r_+ \geq r_-, \Lambda_+ \geq \Lambda_-$ as shown in \ref{Bound on M1}.

\begin{figure}[h]
\begin{center}
\epsfxsize=2in 
\epsfbox{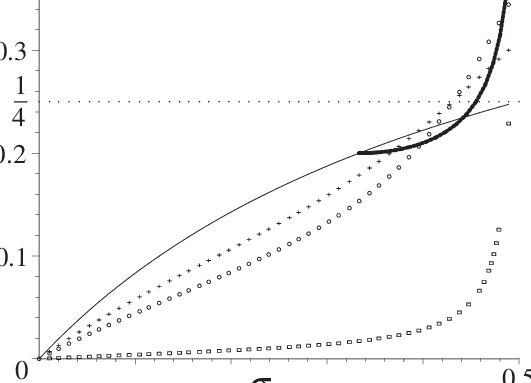}
\hspace{1cm}
\epsfxsize=2in
\epsfbox{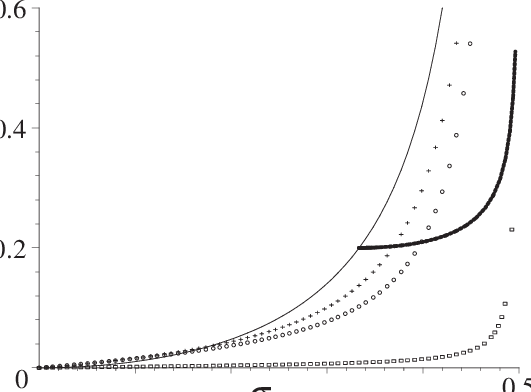}
\end{center}
\caption{\label{Perfect Fluid} The mass per unit length (left) and integral pressure ($P \equiv \mathcal{C}p$, right) for perfect fluid with $\sigma_-=0, \Lambda_-=\Lambda_+ \equiv \Lambda<0, C_-=1$. The plotted curves correspond to $\Lambda=-10,-1,-1/2,-1/1000$ (bottom to top) for a shell of fixed circumference. Pressure always diverges for $\sigma_+ =1/2$. The graph corresponding to $\Lambda=-1/1000$ (full curve) approaches the graph for a LC shell, cf. Fig 3 of \cite{BiZo}. The bold line corresponds to the case $\mu=p$ above which the dominant energy condition is violated for the particular circumference used in the graph.}
\end{figure}

Next we focus on {\em perfect-fluid} interpretation of the induced energy-momentum tensor. If we require $\sigma_-=0$, $\Lambda_-=\Lambda_+ \equiv \Lambda<0$, and $S_{ZZ}=S_{\mathit{\Phi} \mathit{\Phi}}$ and substitute this into (\ref{Induced Tensor with Lambda}) we obtain a simple relation
\begin{equation}
\sinh \sqrt{-3\Lambda} r_+ = \frac {(1-4 \sigma_+^2) \sinh
\sqrt{-3\Lambda} r_-} {4 \sigma_+^2-2 \sigma_+ +1}.
\end{equation}
This enables us to express the density $\mu = S_{TT}$ and pressure $p = S_{ZZ}$ of the fluid as functions of $r_-$ (or of circumference $\mathcal{C}$), $\Lambda$, and $\sigma_+$. For a given $\mathcal{C}$ and $\Lambda$ this amounts to an implicit equation of state given by relations $\mu=\mu(\sigma_+), p=p(\sigma_+)$. For the graphs of $M_1$ and $p$, see \Fref{Perfect Fluid}; here we only point out that $M_1$ can exceed 1/4 unless we impose further conditions, such as the dominant energy condition $\mu \geq p$ for any circumference of the cylinder. The admissible range of the LC parameter reads $\sigma_+ \in [0,1/2)$ (or $\sigma_+ \in [0,1/3]$ with the dominant energy condition).

\subsubsection{Minkowski Inside}
We now turn to the case with flat spacetime inside the cylinder. This corresponds to $\Lambda_-=0$, $\sigma_-=0$, $\Lambda_+<0$, and thus may be considered a `domain wall' in 3+1 dimensions. For perfect fluid, the relation $p = S_{ZZ}=S_{\mathit{\Phi}\mathit{\Phi}}$ (cf. (\ref{Induced Energy-momentum Tensor})) implies
\begin{equation}
\sinh \sqrt{-3\Lambda_+} r_+ = \frac {(1-4 \sigma_+^2) \sqrt{-3\Lambda_+} r_-} {4 \sigma_+^2-2 \sigma_+ +1}.
\end{equation}
Substituting this expression back into (\ref{Induced Tensor with Lambda}), we obtain a relation between density and pressure:
\begin{equation}
\mu + p = \frac {1} {2\pi r_-} \frac {\sigma_+ (1- \sigma_+)} {1- 4 \sigma_+^2},
\end{equation}
which does not depend on $\Lambda_+$ and is thus identical to the Levi-Civita case. The expressions for $M_1 = \mu (2 \pi r_-)$ and the pressure $p$ read
\begin{eqnarray}
M_1 & = & \frac {4 \sigma_+ (1- 2 \sigma_+) +1 - \sqrt {A_+^2-3 \Lambda_+ r_-^2 (1- 4 \sigma_+^2)^2}} {6(1- 4 \sigma_+^2)}, \\
p & = & \frac {2 \sigma_+ (1 + \sigma_+) -1 + \sqrt {A_+^2-3 \Lambda_+ r_-^2 (1- 4 \sigma_+^2)^2}} {12\pi r_-(1- 4 \sigma_+^2)}.
\end{eqnarray}
The numerator of the last expression is always positive ($\Lambda_+<0$) and thus if we require positive and finite pressure we must have $\sigma_+ \in (-1/2,1/2)$. We have $M_1|_{\Lambda_+<0} \leq M_1|_{\Lambda_+=0}=\sigma_+/(1+2\sigma_+)$---the last expression corresponds to the LC case, see \cite{BiZo}. For $M_1$ to be positive we must have $\sigma_+ < -1/2$ or $\sigma_+ \geq 0$. In combination with the restrictions on the pressure we conclude $\sigma_+ \in [0,1/2)$. In this interval of $\sigma_+$, we get $M_1 \leq 1/4$ just like in the Levi-Civita case (however, $M_1$ becomes negative for sufficiently large $\sqrt{\Lambda_+} r_-$).

Let us yet mention solutions with Minkowski inside and azimuthal particles. In this case we obtain
\begin{eqnarray}\label{Phi-motion with Lambda}
M_1 & = & \frac {3 \sigma_+ (1-\sigma_+)} { 2A_+ \cosh (\sqrt {-3\Lambda_+} r_+)-8 \sigma_+^2 +4 \sigma_++1},\\
\label{M-LCL vfi}v_{\Phi}^2 & = & \frac {A_+ \cosh (\sqrt {-3\Lambda_+} r_+)+2 \sigma_+^2 +2 \sigma_+-1} {6 \sigma_+ (1-\sigma_+)}.
\end{eqnarray}
The numerator of (\ref{M-LCL vfi}) is always positive and thus $\sigma_+ \in (0,1)$. Since $v_\Phi^2 \leq 1$ we must further have $\sigma_+ \in [0,1/2]$. Then $ \sigma_+ (1-\sigma_+)/(-8 \sigma_+^2 + 4\sigma_+ +1) \leq M_1|_{\Lambda_+<0} \leq M_1|_{\Lambda_+=0} = \sigma_+ (1-\sigma_+) \leq 1/4$. The bounds on $\sigma_+$ and $M_1$ are just the same as in the Levi-Civita case.

\subsubsection{AdS Inside}
We now consider the anti de Sitter spacetime in horospherical coordinates (\ref{AdS in horospherical coordinates}) inside the cylinder with $\Lambda_- = \Lambda_+ \equiv \Lambda <0$. First fact to mention here is that the circumference at $r=0$ is $2\pi$ and to reach the axis (zero circumference), we need to go farther, to $r \to -\infty$. Therefore, the radial proper distance (measured directly by $r$) from the axis to the surface of the cylinder is always infinite. We thus include within the cylindrical shell a vast part of the AdS space and we can expect the influence of $\Lambda$ to be strong. Indeed, unlike in LC$\Lambda$, there are for example no circular geodesics at constant \emph{horospherical} coordinates $r,z$ in AdS. In addition, we find that we cannot interpret the shell energy-momentum tensor in terms of either massive particles or photons. Perfect fluid requires $\sigma_+=1/2$ necessarily. Equation (\ref{Induced Tensor with Lambda}) implies
\begin{equation}\label{EOS}
\label{pressure} p = \mu \frac {6\pi \mu + \sqrt {-3\Lambda}}
{2(\sqrt {-3\Lambda} -12 \pi \mu)}.
\end{equation}
Inverting the last expression in the form of a series
\begin{equation}
\mu = 2p - 24 \pi \sqrt {-3/\Lambda} p^2 - 1152 \frac {\pi^2}
{\Lambda} p^3 - \ldots,
\end{equation}
we notice that for small pressure, $p = \mu/2$, which corresponds to `2D null fluid'. Notice that $M_1 = 2 \pi \mu \exp \left( \sqrt{-\frac{\Lambda}{3}} \; r_- \right)$ is not bounded for large $r_-$. This is not very surprising as the anti de Sitter metric inside of the shell is the asymptotical form of the metric (\ref{LC+Lambda}) and thus this shell is equivalent to a shell with the same form of metric on both sides of the cut and with $r_- \rightarrow \infty$. This, however, is in sharp contradiction to the assumption $r_+ \geq r_-$ which is a necessary condition for $M_1 \leq 1/4$ (see \ref{Bound on M1}).

\subsection{Black Strings}
Another interesting case is the black-string metric of \cite{Lemos}. This is a typical example where we can explore the physical meaning of the parameters by finding non-singular sources of the spacetime. Let us thus use the AdS spacetime inside our cylindrical shell and the black-string spacetime (\ref{Black-string Metric}) with the same cosmological constant outside. This yields an induced energy-momentum tensor that corresponds to perfect fluid. The shell must necessarily be located above the horizon ($r_+=\sqrt[3]{4M}/\sqrt{-\Lambda/3}$, see \ref{Black-string Metric}). The proper mass per unit length of these cylinders, $M_1 = 2\pi \mu r_+$, reads
\begin{equation}
M_1 = \frac{\alpha r_+}{2} \left( 1- \sqrt{1- \frac{4M}{(\alpha r_+)^3}} \right),
\end{equation}
with $\alpha^2 = -\Lambda/3$. If we let $r_+ \rightarrow \infty$, $M_1$ vanishes while it reaches its limiting value $(M/3)^{1/3}$ for cylinders just above the horizon and we are not able to construct more massive cylinders. As shown in \cite{Lemos}, $M$ is the mass of a cylindrical strip of spacetime defined by unit coordinate length along the $z$-axis which is coordinate dependent and the corresponding proper length vanishes on the axis and diverges at radial infinity. If we define $\widetilde{M}$ as the mass of a section of a unit {\em proper} length, we obtain a different maximum value $M_1 = (2/3)^{1/3} \sqrt{\widetilde{M}}$. In both cases, however, it turns out that $M_1$ is proportional neither to $M$, nor to $\widetilde{M}$, confirming that it measures a different physical property of the solution.
\section{Conclusions} We considered static, vacuum, cylindrically symmetric spacetimes with a non-zero cosmological constant, their geometrical properties, and their sources formed by cylindrical shells of counter-rotating photons or dust and of perfect fluid. We have shown that even with $\Lambda \not =0$, the LC parameter must be confined to the interval $\sigma \in [0,1/2]$, i.e., the same range as for $\Lambda=0$. To the first order, it determines the mass $M_1$ per unit length of the source. The range of $M_1$ is no longer bounded by 1/4. This fact thus represents the main difference from the Levi-Civita case.
\section*{Acknowledgments} This work was supported by grants GA\v{C}R 202/06/0041, GA\v{C}R 202/05/P127, by the Centre for Theoretical Astrophysics, and by research project MSM 0021620860. The authors acknowledge the kind hospitality of the Max-Planck Society and Albert-Einstein Institute in Golm, Germany. J.B. would also like to thank the Alexander von Humboldt Foundation.
\appendix
\section{Derivation of the Metric} \label{Derivation of the Metric}
Following \cite{Tian} and denoting the proper radius by $r$, we may write a static cylindrically symmetric metric as
\begin{equation}\label{Metric-Definition}
ds^2 = -\exp A(r) \; dt^2 + dr^2 + \exp B(r) \; dz^2 + \exp C(r) \;d\varphi^2, \label{Line Element}
\end{equation}
here $t,z \in \! I\hspace{-0.13cm}R,r \in \! I\hspace{-0.13cm}R^+, \varphi \in [0,2\pi)$. Solving Einstein equations in vacuum with a non-zero cosmological constant $\Lambda$, we get
\begin{equation}\label{Equation for Sigma}
\frac {d\:^2 \Sigma(r)} {dr^2} + \frac {1}{2} \left( \frac {d\Sigma(r)} {dr} \right)^2 + 6 \Lambda = 0,
\end{equation}
where $\Sigma(r) = A(r) + B(r) + C(r)$. The solution for $\Lambda>0$ reads
\begin{equation}
\Sigma(r) = const + 2 \ln \sin \left(\sqrt{3 \Lambda} (r+R) \right)
\end{equation}
with $R$ the other integration constant. In the following, we only need its first derivative (there are two solutions: the other solution is shifted along $r$ which can always be incorporated into $R$)
\begin{equation}
\frac {d \Sigma(r)} {dr} = 2 \sqrt{3 \Lambda} \cot \left(\sqrt{3 \Lambda} (r-R) \right).
\end{equation}
For negative $\Lambda$, we find
\begin{eqnarray}
\frac {d \Sigma(r)} {dr} = 2 \sqrt{-3 \Lambda} \coth \left(\sqrt{-3 \Lambda} (r-R) \right), \label{Sol for negative Lambda}\\
\frac {d \Sigma(r)} {dr} = 2 \sqrt{-3 \Lambda} \tanh \left(\sqrt{-3 \Lambda} (r-R) \right), \label{The other sol for negative Lambda}\\
\frac {d \Sigma(r)} {dr} = \pm 2 \sqrt{-3 \Lambda}. \label{Anti de Sitter}
\end{eqnarray}
Einstein equations further imply
\begin{equation}
\frac {d\:^2 f(r)} {dr^2} + \frac {1}{2} \frac {d \Sigma(r)}{dr} \frac {df(r)} {dr} + 2 \Lambda =0
\end{equation}
for each of the functions $A(r), B(r),$ and $C(r)$. For $\Lambda>0$, we can thus write
\begin{equation}
\exp(f(r)) = \beta \tan^\alpha x \; \sin^{2/3} 2x,
\end{equation}
where $x=\sqrt{3 \Lambda}(r+R)/2$ and $\alpha, \beta$ are integration constants. By rescaling the coordinates we can get rid of $\beta$ (however, this changes the range of $\varphi$ and we must introduce the conicity parameter $C$, which also takes care of the units of $g_{\varphi \varphi}$). If we now insert the solutions back into Einstein equations, we obtain conditions on the three constants $\alpha$:
\begin{eqnarray}
0 & = & \alpha_A+\alpha_B+\alpha_C, \nonumber \\
-4/3 & = & \alpha_A \alpha_B + \alpha_A \alpha_C + \alpha_C \alpha_B. \label{Algebraic relations}
\end{eqnarray}
Now let us introduce the following parametrization by the Levi-Civita parameter\footnote{The Levi-Civita parameter is denoted by $\sigma$ in the present paper. Its relation to the standard value $\sigma=m/2$ follows from the transformation to the Levi-Civita metric in the limit $\Lambda \rightarrow 0$; see, e.g., \cite{da_Silva_et_al}.}
\begin{eqnarray*}
\alpha_A & = & 4\sigma/A - 2/3,\\
\alpha_B & = & 4\sigma(2 \sigma - 1)/A - 2/3,\\
\alpha_C & = & 2(1-2\sigma)/A - 2/3,
\end{eqnarray*}
with
\begin{equation}
A \equiv 4\sigma^2-2\sigma+1.
\end{equation}
This yields (\ref{LC+Lambda}) (we put $R=0$ to place the axis at $r=0$ for $\sigma<1/2$---but we can as well place the axis somewhere else). We obtain an analogous solution for $\Lambda<0$ by following the same lines with Eqs. (\ref{Sol for negative Lambda}) and (\ref{The other sol for  negative Lambda}). However, we do not obtain a real solution from (\ref{The other sol for  negative Lambda}) (only complex solutions since in Eq. (\ref{Algebraic relations}) we now have a positive LHS) and the only possible metric has $P(r)=2\tanh (\sqrt{-3\Lambda}r/2)/\sqrt{3\Lambda}, \; Q(r)= \sinh (\sqrt{-3\Lambda}r)/\sqrt{3\Lambda}$. We thus only need to substitute the trigonometric functions by their hyperbolic counterparts and $\Lambda \rightarrow |\Lambda|$---see \cite{da_Silva_et_al}. $C$ is the conicity parameter related to the missing angle that enables us to restore the range of $\varphi$ to $[0,2\pi)$---see Eq. (\ref{Conicity definition}).

Equation (\ref{Anti de Sitter}) yields another independent solution for negative $\Lambda$, namely
\begin{equation}\label{AdS in horospherical coordinates}
ds^2 = dr^2 + \exp \left( \pm 2\sqrt{-\frac{\Lambda}{3}} \; r \right) \left[dz^2 + d\varphi^2 - dt^2 \right],
\end{equation}
where $\varphi$ has been rescaled from $[0,2\pi)$ to a new interval and $r \in [0,\infty)$. As $r=0$ represents no special hypersurface, we can also extend the range of $r$ to $(-\infty,\infty)$ and write the metric with the `$+$' sign only. Spacetime with metric (\ref{AdS in horospherical coordinates}) constitutes a part of the anti de Sitter spacetime with metric written in the horospherical coordinates. This follows from the transformation
\begin{equation}
\begin{array}{lcl}
r = \mp \sqrt{-\frac {3}{\Lambda}} \ln (x \sqrt{\frac {-\Lambda}{3}}) & \mbox{or} & x = \sqrt{-\frac {3}{\Lambda}} \exp \left( \mp r\sqrt{\frac {-\Lambda}{3}} \right)\\
\exp \left( \pm 2\sqrt{-\frac{\Lambda}{3}} \; r \right) = -\frac {3}{\Lambda x^2} & \Rightarrow & dr = \mp \sqrt{-\frac {3}{\Lambda}} \frac {dx}{x},\\
\varphi \in [0,a) \rightarrow y \in \! I\hspace{-0.13cm}R,
\end{array}
\end{equation}
which gives the conformally flat form of the AdS metric
\begin{equation}\label{Conformal AdS}
ds^2 = \frac {3}{(-\Lambda) x^2} \left[ -dt^2 + dx^2 + dy^2 +dz^2 \right].
\end{equation}
Solution (\ref{AdS in horospherical coordinates}) with the `$+$' sign covers that part of the anti-de Sitter hyperboloid where $x \in (0,\sqrt{-3/\Lambda}]$ while the `$-$' solution is valid for $x \in [\sqrt{-3/\Lambda},\infty)$. However, the axis of (\ref{AdS in horospherical coordinates}) is not located at $r=0$ (circumference $\mathcal{C}=2\pi$)---rather, it is at $r \rightarrow \infty$ for the `$-$' sign which corresponds to $x \rightarrow \infty$ and it is not present in the spacetime at all for the `$+$' sign as follows from the expression for the circumference $\mathcal{C} = 2\pi \exp(\pm r \sqrt {-\Lambda/3}) =  2\pi \sqrt{-3/\Lambda} /x$. Finally, we had to extend the range of $\varphi$ to $I\hspace{-0.13cm}R$ otherwise we would only have a strip of the resulting spacetime. To cover the entire anti de Sitter manifold, we need 4 such coordinate systems with $r \in [0,\infty)$ or 2 systems with $r \in (-\infty,\infty)$.
\section{Asymptotic comparison of Schwarzschild-anti de Sitter and LC$\mathbf{\Lambda}$ with $\mathbf{\Lambda<0}$}\label{Asymptotics}

We first give the transformation of anti de Sitter metric from Schwarzschild coordinates $(T,R,\Theta,\Phi)$ to horospherical coordinates $(t,r,z,\varphi)$
\begin{eqnarray}\label{Schwarzschild to horospherical}
\fl \hspace{1.5cm} T = \alpha \arctan \frac {\alpha^2 - t^2 + \varphi^2 + z^2 + \alpha^2 \exp (\pm 2 r / \alpha)} {2\alpha t}, \nonumber\\
\fl \hspace{1.5cm} R = \frac {\exp(\mp r/\alpha)} {\alpha} \sqrt{\alpha^2 (\varphi^2 + z^2) + [ \alpha^2 + t^2 - \varphi^2 - z^2 - \alpha^2 \exp (\pm 2 r / \alpha) ]^2/4}, \nonumber\\
\fl \hspace{1.5cm} \Theta = \arctan \frac{2\alpha \sqrt{\varphi^2+ z^2}}{\alpha^2 + t^2 - \varphi^2 - z^2 - \alpha^2 \exp (\pm 2 r / \alpha)}, \nonumber\\
\fl \hspace{1.5cm} \Phi = \arctan \frac {z} {\varphi},
\end{eqnarray}
where $\alpha^2=3/(-\Lambda)$. The inverse transformation reads
\begin{eqnarray}\label{Horospherical to Schwarzschild}
\fl \hspace{1.5cm} t = \frac {\alpha \sqrt{R^2 + \alpha^2} \: \cos \frac {T}{\alpha}} {R\cos \Theta + \sqrt{R^2 + \alpha^2} \: \sin \frac {T}{\alpha}}, \nonumber\\
\fl \hspace{1.5cm} r = \pm \alpha \ln \frac {\alpha} {R\cos \Theta + \sqrt{R^2 + \alpha^2} \: \sin \frac {T}{\alpha}}, \nonumber\\
\fl \hspace{1.5cm} z = \frac {\alpha R \sin \Theta \sin \Phi} {R\cos \Theta + \sqrt{R^2 + \alpha^2} \: \sin \frac {T}{\alpha}}, \nonumber\\
\fl \hspace{1.5cm} \varphi = \frac {\alpha R \sin \Theta \cos \Phi} {R\cos \Theta + \sqrt{R^2 + \alpha^2} \: \sin \frac {T}{\alpha}}.
\end{eqnarray}
Applying this transformation to $ds^2 = -(1+ r^2/\alpha^2) \; dt^2 + dr^2 / (1+ r^2/\alpha^2)+ r^2 \; d\Omega^2$ yields metric (\ref{AdS in horospherical coordinates}). We now take Schwarzschild-anti de Sitter metric $ds^2 = -(1-2M/r+ r^2/\alpha^2) \; dt^2 + dr^2 / (1-2M/r+ r^2/\alpha^2) + r^2 \; d\Omega^2$ and transform it using (\ref{Schwarzschild to horospherical}) with the lower signs. Asymptotically, we have for $g_{\mu\nu}$ in $[t,r,z,\varphi]$
\begin{equation*}
\fl \left(\begin{array}{c|c|}
-\mbox{e}^{(2\frac{r}{\alpha})} + \mbox{e}^{(-\frac{r}{\alpha})} M (\ldots) + M \left[ \ldots \right] & M t \mbox{e}^{(-3\frac{r}{\alpha})} (\ldots) + M t \left[ \ldots \right] \\
.& 1 + M \mbox{e}^{(-3\frac{r}{\alpha})} (\ldots) + M \left[ \ldots \right] \\
.&.\\
.&.\\
\end{array}
\right.\end{equation*}
\begin{equation}\label{Asymptotics of SAdS}
\fl \hspace{0.5cm} \left.\begin{array}{c|c}
M t z \mbox{e}^{(-\frac{r}{\alpha})} (\ldots) + M t z \left[ \ldots \right] & M t \varphi \mbox{e}^{(-\frac{r}{\alpha})} (\ldots) + M t \varphi \left[ \ldots \right]\\
M z \mbox{e}^{(-3\frac{r}{\alpha})} (\ldots) + M z \left[ \ldots \right] & M \varphi \mbox{e}^{(-3\frac{r}{\alpha})} (\ldots) + M \varphi \left[ \ldots \right] \\
\mbox{e}^{(2\frac{r}{\alpha})} + \mbox{e}^{(-\frac{r}{\alpha})} M z^2 (\ldots) + M z^2 \left[ \ldots \right] & M z \varphi \mbox{e}^{(-\frac{r}{\alpha})} (\ldots) + M z \varphi \left[ \ldots \right] \\
.& \mbox{e}^{(2\frac{r}{\alpha})} + \mbox{e}^{(-\frac{r}{\alpha})} M \varphi^2 (\ldots) + M \varphi^2 \left[ \ldots \right]\\
\end{array}
\right)\end{equation}
%
%
\rem{
\begin{equation}
\fl \left(\begin{array}{c|c|c|c}
-\mbox{e}^{(2\frac{r}{\alpha})} + M \left[ \mbox{e}^{(-\frac{r}{\alpha})} (\ldots) + \ldots \right] & M t \left[ \mbox{e}^{(-3\frac{r}{\alpha})} (\ldots) + \ldots \right] & M t z \left[ \mbox{e}^{(-\frac{r}{\alpha})} (\ldots) + \ldots \right] & M t \varphi \left[ \mbox{e}^{(-\frac{r}{\alpha})} (\ldots) + \ldots \right]\\
& 1 + M \left[ \mbox{e}^{(-3\frac{r}{\alpha})} (\ldots) + \ldots \right] & M z \left[ \mbox{e}^{(-3\frac{r}{\alpha})} (\ldots) + \ldots \right] & M \varphi \left[ \mbox{e}^{(-3\frac{r}{\alpha})} (\ldots) + \ldots \right] \\
&& \mbox{e}^{(2\frac{r}{\alpha})} + M z^2 \left[ \mbox{e}^{(-\frac{r}{\alpha})} (\ldots) + \ldots \right] & M z \varphi \left[ \mbox{e}^{(-\frac{r}{\alpha})} (\ldots) + \ldots \right] \\
&&& \mbox{e}^{(2\frac{r}{\alpha})} + M \varphi^2 \left[ \mbox{e}^{(-\frac{r}{\alpha})} (\ldots) + \ldots \right] \\
\end{array}
\right)\end{equation}}
The LC$\Lambda$ metric gives
\begin{equation*}
\fl \left(
\begin{array}{c|c|}
-\mbox{e}^{(2\frac{r}{\alpha})} + \mbox{e}^{(-\frac{r}{\alpha})}(-4/3 + 8\sigma) + \left[ \ldots \right] & 0\\
.& 1\\
.&.\\
.&.\\
\end{array}
\right.
\end{equation*}
\begin{equation}\label{Asymptotics of LCC}
\fl \hspace{0.5cm} \left.
\begin{array}{c|c}
0&0\\
0&0\\
\mbox{e}^{(2\frac{r}{\alpha})} + \mbox{e}^{(-\frac{r}{\alpha})}(4/3 + 8\sigma) + \left[ \ldots \right] &0\\
.& \mbox{e}^{(2\frac{r}{\alpha})} + \mbox{e}^{(-\frac{r}{\alpha})}(-8/3 + 16\sigma^2) + \left[ \ldots \right] \\
\end{array}
\right)
\end{equation}
%
%
\rem{\begin{equation}
\fl \left(\begin{array}{c|c|c|c}
-\mbox{e}^{(2\frac{r}{\alpha})} + \mbox{e}^{(-\frac{r}{\alpha})}(-4/3 + 8\sigma) + \ldots &&&\\
& 1&&\\
&& \mbox{e}^{(2\frac{r}{\alpha})} + \mbox{e}^{(-\frac{r}{\alpha})}(4/3 + 8\sigma) + \ldots &\\
&&& \mbox{e}^{(2\frac{r}{\alpha})} + \mbox{e}^{(-\frac{r}{\alpha})}(-8/3 + 16\sigma^2) + \ldots\\
\end{array}
\right)\end{equation}}
The radial dependence of the individual metric components expanded in a series thus coincides up to $\mbox{e}^{(-\frac{r}{\alpha})}$. For discussion, see pages \pageref{Discussion of asymptotics} and \pageref{Discussion of asymptotics  part II}.
\section{Bound on $\mathbf{M_1}$}\label{Bound on M1}
In our previous work, we proved that $M_1 \leq 1/4$ in all cases with Minkowski inside and a general LC outside, cf. Eq. (6) in \cite{BiZo}. For $\Lambda \not= 0$, we presented in the main text above several examples with an upper bound on $M_1$, e.g., if the shell is composed of counter-rotating particles and we further assume $\Lambda_-=\Lambda_+ \equiv \Lambda<0$ and put $\sigma_-=0, C_-=1$ so that there is no singularity within the entire spacetime, see \Fref{Azimuthal Particles}. Here, we generalize this result without assuming a specific form of $T_{\mu\nu}$ and admitting even different values of $\Lambda_-$ and $\Lambda_+$. We only assume $\Lambda_- \leq \Lambda_+$ and $r_- \leq r_+$.

We proceed from the general form of the induced energy-momentum tensor with $\sigma_-=0, \mathcal{C}_-=1$ (cf. (\ref{Induced Tensor with Lambda}))
\begin{eqnarray}
\fl S_{TT} & = & \frac{ \sqrt{-\Lambda_+/3 }} {8\pi \sinh (\sqrt {-3\Lambda_+} r_+)} \left[-2\cosh (\sqrt {-3\Lambda_+} r_+) + \frac{  -4\sigma_+^2 +8\sigma_+ -1} {4\sigma_+^2 -2\sigma_+ +1} \right] + \nonumber \\
\fl && + \frac{ \sqrt{-\Lambda_-/3 }} {8\pi} \frac {\left(2\cosh (\sqrt {-3\Lambda_-} r_-) +1 \right)} {\sinh (\sqrt {-3\Lambda_-} r_-) }.
\end{eqnarray}
Circumference ${\mathcal{C}}$ is given by ${2\pi \sqrt{g_{\varphi\varphi}(\Lambda_-,r_-)}}$ and ${M_1 = \mathcal{C} (\Lambda_-,r_-) \, S_{TT}(\sigma_+,\Lambda_\pm,r_\pm)}$ is thus a function of 5 independent parameters. Evaluating $dM_1/d\sigma_+$, we find the extremum at $\sigma_+=1/2$. We can thus write
\begin{eqnarray}\label{Sigma=1/2}
\fl S_{TT}|_{\sigma_+=1/2} & = & \frac{ \sqrt{-\Lambda_+/3 }} {8\pi} \frac {\left(-2\cosh (\sqrt {-3\Lambda_+} r_+) +2 \right)} {\sinh (\sqrt {-3\Lambda_+} r_+) } + \nonumber \\
\fl && + \frac{ \sqrt{-\Lambda_-/3 }} {8\pi} \frac {\left(2\cosh (\sqrt {-3\Lambda_-} r_-) +1 \right)} {\sinh (\sqrt {-3\Lambda_-} r_-) }.
\end{eqnarray}
This function reaches its maximum value at $r_+=0$:
\begin{equation}\label{r_+=0}
S_{TT}|_{r_+=0} = \frac{ \sqrt{-\Lambda_-/3 }} {8\pi} \frac {\left(2\cosh (\sqrt {-3\Lambda_-} r_-) +1 \right)} {\sinh (\sqrt {-3\Lambda_-} r_-) }.
\end{equation}
Introducing a new variable $X = \sqrt {-3 \Lambda_-}r_-/2$, we find
\begin{equation}
M_1 = \frac{(2\cosh 2X+1)\sinh X} {6\cosh ^{1/3} (X)\sinh 2X} = \frac{4\cosh^2 X-1} {12\cosh ^{4/3} X}.
\end{equation}
This is \emph{not} bounded. However, we already found a physically plausible condition limiting the range of $r_+$, namely $r_+ \geq r_-$ (see the discussion below (\ref{particle_condition})). Hence, instead of maximizing (\ref{Sigma=1/2}) by $r_+=0$, we set $r_+=r_-$. Analogously to the situation with $r_+$ and $r_-$, we find that for $\Lambda_+$ and $\Lambda_-$ completely independent, $M_1$ can diverge again. However, assuming $r_+ \geq r_-$ and, in addition, $\Lambda_+ \geq \Lambda_-$, we finally arrive at
\begin{equation}
M_1 = \frac{1} {4} \frac {1} {\cosh^{4/3}( \sqrt {-3\Lambda_-}r_-/2)
}.
\end{equation}
Thus for any values of $r_-$ and $\Lambda_-$ we always have $M_1 \leq 1/4$. It can be shown that the bound $M_1 \leq 1/4$ applies to the case $\Lambda_->0$ and $\Lambda_+<0$ as well.

Let us summarize: With a vanishing cosmological constant, $\Lambda_+=\Lambda_-=0$, the bound $M_1 \leq 1/4$ holds always if the entire spacetime is regular as proved in \cite{BiZo}. Here we have shown that admitting a non-zero cosmological constant, we have to require not only a regular spacetime but also additional conditions, such as $r_+ \geq r_-$ and $\Lambda_+ \geq \Lambda_-$.

\section*{References}\label{References}
\addtocontents{toc}{\contentsline {section}{\numberline {}\hspace{-0.75cm} References}{\pageref{References}}} 


\begin{thebibliography}{99}
    \bibitem {BiZo} Bi\v{c}\'ak J and \v{Z}ofka M 2002 \CQG {\bf 19} 3653
    \bibitem {Israel} Israel W 1966 \NC {\bf  B 44} 1 (erratum {\bf B 49} 463)    
    \bibitem {Domain_walls} Peebles P J E 1993 {\it Principles of Physical Cosmology} (Princeton University Press)
    \bibitem {Bondi} Bondi H 2004 {\it Proc. R. Soc. Lond. A} {\bf 460} 463
    \bibitem {Linet} Linet B 1986 \JMP {\bf 27} 1817
    \bibitem {Tian} Tian Q 1986 \PR D {\bf 33} 3549
    \bibitem {Kramer} Stephani H, Kramer D, MacCallum M, Hoenselaers C and Herlt E 2003 {\it Exact Solutions to Einstein's Field Equations (2nd ed.)} (Cambridge University Press)
    \bibitem {Lemos} Lemos J P S 1995 \CQG {\bf 12} 1081; Phys. Lett. B {\bf 353} 46
    \bibitem {Horowitz} Horowitz G 1993 {\it The Dark Side of String Theory: Black Holes and Black Strings} Trieste Spring School on String Theory and Quantum Gravity 1992
    \bibitem {da_Silva_et_al} da Silva M F A, Wang A Z, Paiva F M and Santos N O 2000 \PR D {\bf 61} 044003
    \bibitem {Brown_and_York} Brown J D and York Jr. J W 1993 \PR D {\bf 47} 1407
    \bibitem {Philbin} Philbin T G 1996 \CQG {\bf 13} 1217
\end{thebibliography}
\end{document}